\documentclass{aa}  
\usepackage{graphicx}
\usepackage{txfonts}
\usepackage[colorlinks]{hyperref}
\usepackage[usenames]{xcolor}
\usepackage{siunitx}
\usepackage{multirow}
\usepackage[utf8]{inputenc}
\usepackage{alphabeta}

\newcommand{\E}[1]{$\times10^#1$}
\newcommand{\percent}{\%}

\DeclareSIUnit\photon{hν}
\DeclareSIUnit\electron{e^-}
\DeclareSIUnit\adu{ADU}
\DeclareSIUnit\Jansky{Jy}

\begin{document}

\title{AGILIS:\\Agile Guided Interferometer for Longbaseline Imaging Synthesis}

\subtitle{Demonstration and concepts\\of reconfigurable optical imaging interferometers}

\author{Julien Woillez\inst{1,2}
\and Olivier Lai\inst{3,4,5,6} 
\and Guy Perrin\inst{7} 
\and François Reynaud\inst{8} 
\and Marc Baril\inst{3} 
\and Yue Dong\inst{9} 
\and Pierre Fédou\inst{7} } 

\institute{European Southern Observatory, Karl-Schwarzschild-Str. 2, Garching bei M\"{u}nchen, 85748, Germany.
\and W. M. Keck Observatory, 65-1288 Mamalahoa Hwy., Kamuela, HI 96743, USA.
\and Canada France Hawaii Telescope, 65-1238 Mamalahoa Hwy., Kamuela, HI 96743, USA.
\and Gemini Observatory, 670 A'ohoku Place, Hilo, HI 96720, USA.
\and Subaru Telescope, 650 A'ohoku Place, Hilo 96720, USA.
\and Université Côte d'Azur, Observatoire de la Côte d'Azur, CNRS, Laboratoire Lagrange, Bd de l'Observatoire, CS 34229, 06304 Nice cedex 4, France.
\and LESIA, Observatoire de Paris, CNRS, UPMC, Universit\'e Paris-Diderot, Paris Sciences et Lettres Research University, 5 place Jules Janssen, 92195 Meudon, France.
\and XLIM, 123 avenue Albert Thomas, 87060 Limoges, France.
\and Ecole Polytechnique, Laboratoire de Physique des Plasmas, Route de Saclay, 91128 Palaiseau, France.}

\date{Received January 26, 2017; accepted March 6, 2017}

\abstract
{In comparison to the radio and sub-millimetric domains, imaging with optical interferometry is still in its infancy. Due to the limited number of telescopes in existing arrays, image generation is a demanding process that relies on time-consuming reconfiguration of the interferometer array and super-synthesis.}
{Using single mode optical fibres for the coherent transport of light from the collecting telescopes to the focal plane, a new generation of interferometers optimized for imaging can be designed.}
{To support this claim, we report on the successful completion of the `OHANA Iki project: an end-to-end, on-sky demonstration of a two-telescope interferometer, built around near-infrared single mode fibres, carried out as part of the `OHANA project.}
{Having demonstrated that coherent transport by single-mode fibres is feasible, we explore the concepts, performances, and limitations of a new imaging facility with single mode fibres at its heart: Agile Guided Interferometer for Longbaseline Imaging Synthesis (AGILIS).}
{AGILIS has the potential of becoming a next generation facility or a precursor to a much larger project like the Planet Formation Imager (PFI).}


\maketitle

\section{Introduction}
\label{sec:intro}
Optical stellar interferometry has flourished in recent years with the advent of large telescope arrays and imaging arrays, leading to important astrophysical results being achieved \citep[e.g.][]{Tuthill+1999, Monnier+2004, Kraus+2005, Monnier+2007, Kloppenborg+2010}.
The combination of large aperture and long baselines has improved our understanding of many types of stars and has opened up fields outside of stellar physics \citep{Swain+2003, Jaffe+2004, Tristram+2009}.
Even though much progress has been made in the field of image synthesis, the vast majority of astrophysics carried out with existing interferometers is still model based.
This is not surprising if one considers the large number of apertures, or alternatively, the large amount of time taken using fewer telescopes, that is required to map out the (u,~v) plane and produce images of sufficient resolution to reach specific science goals.

In most existing interferometers, the telescopes are set at fixed positions in the ground; this ensures a stable baseline.
However, only a few discrete spatial frequencies can be sampled at any given time, therefore only some portions of the (u,~v) plane can be mapped depending on which part of the sky is being observed.
Some interferometers such as IOTA or the VLTI's ATs can be repositioned at discrete predetermined positions.
Although this improves the (u,~v) plane coverage, changing the telescopes positions usually relies on heavy infrastructure, requires very accurate mechanical positioning systems, and still delivers a limited (u,~v) coverage that compromises the imaging performance. 

In the foreseeable future, imaging with milli-arcsecond resolution in the optical/IR domain will be done most efficiently using sparse or diluted pupil interferometry.
Still, restricted (u,~v) plane coverage will limit the quality of image reconstruction, and the dynamic range of the optical delays may restrict sky coverage for the longest baselines.

We have recently demonstrated a deployable interferometer that uses single-mode fibres for coherent light transport in the context of the `OHANA (Optical Hawaiian Array for Nanoradian Astronomy) project (Section~\ref{Sec:OhanaIki}).
In this paper, we leverage off this experience to propose a new concept for an imaging interferometer that will achieve a high spatial frequency (u,~v) plane coverage (Section~\ref{sec:Agilis}).

\section{On-sky demonstration of the reconfigurable optical interferometer `OHANA Iki}\label{Sec:OhanaIki}

At a time when the development of adaptive optics (AO) on large telescopes was expected to improve the sensitivity of optical interferometers \citep{Mariotti1994}, and the use of single-mode fibres in the beam combiner of an optical interferometer had just been demonstrated with the FLUOR instrument \citep{CoudeDuForesto&Ridgway1992,Perrin+1998}, \citet{Mariotti+1996} proposed the `OHANA (Optical Hawaiian Array for Nano-radian Astronomy) concept: turning the large AO-equipped telescopes on the summit of Mauna Kea into a sensitive optical interferometer using coherent beam transport with single-mode fibres.

Around 2000, with the approval of the Mauna Kea telescopes' governing bodies, the `OHANA concept moved towards implementation \citep{Woillez+2001}.
Coupling between AO systems and single-mode fibres in the J, H, and K bands was tested on the Canada-France-Hawaii (CFH) telescope and the Keck II telescope in 2002, and on the Gemini telescope in 2003 \citep{Woillez+2004}.
The development and characterisation of a pair of \SI{300}{\meter} long, single-mode fluoride glass fibres for the K band \citep{Kotani+2005} made the bypass of the Keck Interferometer \citep[KI,][]{Colavita+2013} beamtrains possible.
The first on-sky fringes were demonstrated in 2005, with the coherent light transport achieved in single-mode fibres between the two AO Nasmyth foci and the interferometric laboratory \citep{Perrin+2006a}.

In parallel to this initial demonstration on the already operational KI, the project embarked on the development of a delay line, a beam combiner, and additional pairs of single-mode fibres for the J and H bands \citep{Vergnole+2004,Vergnole+2005}, aiming for a second demonstration on the \SI{160}{\meter} baseline between the two different telescopes: CFHT and Gemini.
In order to test the functionalities of the hardware, before attempting to link two large Mauna Kea telescopes, it was deemed prudent to demonstrate routine interference fringes using small, commercially available telescopes, validating the entire interferometric setup independently.
This is how the `OHANA Iki\footnote{Iki means small in the Hawaiian language.} project came to life.

\subsection{Instrument description}\label{SSec:InstrumentDescription}

The `OHANA Iki instrument is made of two Iki telescopes (section \ref{SSSec:IkiTelescopes}), connected with two pairs of \SI{300}{\meter} long J and H band polarisation maintaining fibres to a delay-line (section \ref{SSSec:DelayLine}) and a beam combiner (section \ref{SSSec:BeamCombiner}).
All components, including the capture of detector frames, is controlled through a single graphical user interface, or GUI (section \ref{SSSec:ControlSoftware}).
The layout of the instrument inside the CFHT coudé room and outside the observatory facility is illustrated in Fig.~\ref{Fig:OhanaIkiLayout}.

\subsubsection{Iki telescopes}\label{SSSec:IkiTelescopes}

A detailed description of the `OHANA Iki hardware can be found in \citet{Baril+2010}.
Briefly, it consists of two independent units that can be set up a variable distance apart.
Each unit consists of a \SI{203.2}{\milli\meter} Schmidt Cassegrain Telescope (SCT, Celestron CPC 800), injection optics, and a control host computer.

After being folded by a tip-tilt mirror at the Cassegrain focus, the incoming light is separated using a beam-splitter cube.
Half of the light is directed to a camera (Prosilica GE 680) used for acquisition in full field or for fast guiding in windowed mode.
The other half is refocused onto the fibre using an achromat to match the fibre input F-ratio.
The total throughput in J band (\SI{1.25}{\micro\meter}) is not ideal; half of the light is lost at the beam splitter ($T=50\percent$), the Schmidt corrector plate is not optimized at near infrared wavelengths ($T=70\percent$), and the telescope central obscuration of $31\percent$ in diameter reduces the theoretical coupling into the nominally gaussian, fundamental fibre mode ($T=64.7\percent$).
Nonetheless, this $22.6\percent$ non-turbulent transmission was deemed sufficient for a proof-of-concept demonstration. 

The tip-tilt is a voice-coil driven \SI{25}{\milli\meter} diameter mirror (Optics-in-Motion OIM-101 unit).
The camera full-frame covers a rectangular field of view of \SI{8}{\arcmin} by \SI{6}{\arcmin}.
When reading a sub-window, guiding frequencies up to \SI{500}{\hertz} are possible.
In this fast-guiding mode, the correction bandwidth (\SI{0}{\deci\bel} crossing of the attenuation curve) was measured at about \SI{40}{\hertz}, and is limited by the dynamical properties of the tip-tilt mirror rather than the servo loop rate.

Once the telescope is manually aligned with $2 \sim 3$ bright stars in the viewfinder, any subsequent control can be done remotely.
On request, the telescope slews to a target; once tracking, a spiral search is performed until the brightest target is detected in the camera's full frame.
The control software then moves the star toward a predefined pixel coordinate that corresponds to the fibre core location; at this point fast guiding is enabled to ensure stable injection of starlight into the fibre.
In addition to star acquisition requests, the `OHANA control software interacts with the host computers to perform raster-scans used to optimize injection efficiency into the fibre (see Fig.~\ref{Fig:RasterScan}).
The units are monitored remotely via VNC using a real-time GUI to obtain feedback on the image quality and guiding stability.

The telescopes routinely achieve \SIrange{0.1}{0.2}{\arcsec} rms image stability depending on the seeing and wind; turbulent airflow at the ground affects injection stability.
With the focal reducer, the \SI{5}{\micro\meter} fibre subtends \SI{1.5}{\arcsec} on the sky, so a stability of  1/10$^{th}$ core should be achieved.

The telescopes can be carried or rolled outside and set up in about 15 minutes.
They are preferably set up on the soft cinder away from the observatory building.
The telescopes' relative separation and orientation are then roughly adjusted by using a ruler and taking a bearing on landmarks.
We estimate a relative separation accuracy on the order of \SI{1}{\centi\meter}, and an orientation accuracy of the resulting baseline on the order of \SI{5}{\degree}.
Once each telescope is set-up and stabilized on the target star, the starlight is sent through the fibres to the delay line and beam combiner.

\subsubsection{Delay line}\label{SSSec:DelayLine}

The delay line was designed for the \SI{163}{\meter} baseline between the CFH and Gemini telescopes.
With the CFH Coudé room being the only readily available installation location, its physical extent was limited to \SI{15}{\meter}.
In order to achieve a reasonable sky coverage, the optical path had to be folded multiple times.
As illustrated in Fig.~\ref{Fig:DelayLine}, a \SI{\pm50}{\meter} optical path stroke is realized with a mobile carriage on \SI{12.5}{\meter} long rails used from both sides ($\times 2$) and in quadruple pass ($\times4$).
The central carriage can be relocated between two observations in order to adjust the general pointing of the interferometer.
While observing, an additional \SI{1}{\meter} long translation stage (Aerotech ALS50100) in double pass provides the \SI{2}{\meter} continuous delay capability required to track the sidereal fringe motion up to \SI{5}{\milli\meter\per\second}.
This \SI{\pm50}{\meter} delay line on a quasi North-South \SI{163}{\meter} long CFH Gemini baseline, provides a close-to-transit pointing capability for all targets within a declination span of $\delta=$ \SI{19.8}{\degree}\SI{\pm18}{\degree}.
The sidereal velocity of the fringes, for targets at a zenith angle smaller than \SI{45}{\degree}, is smaller than \SI{3}{\milli\meter\per\second}.
The resulting continuous sidereal tracking autonomy is therefore always longer than \SI{10}{\minute}.
In order to minimize the down time between two continuous sidereal tracking intervals, the central carriage motion is automatically controlled using a motorized belt and a linear encoder.
The optical coupling with the single-mode fibres is achieved using off-axis parabolas.
Dihedral mirrors are used to implement the multiple-pass beam folding.
Pitch variations of the central carriage, due to the alignment errors of the rails, are corrected with a pitch actuator on that same carriage, using an empirically determined lookup table.
Finally, following a relocation of the central carriage, the final fibre-to-fibre alignment is achieved with fine steering mirrors.

\subsubsection{Beam combiner}\label{SSSec:BeamCombiner}

The multi-coaxial beam combiner originally anticipated for `OHANA has been presented in \citet{Woillez2004}.
However, when used in the `OHANA Iki setup, this beam combiner is simplified to a fibred X coupler, the outputs of which are re-imaged on an infrared detector after their polarizations have been separated using a Wollaston prism.
This coupler is located on an optical bench inside the CFH Coudé room next to the delay line, and was used for all the results reported here.
Fringe contrast measurements are performed with an OPD scanner integrated into the delay line.
Specifically, a flat mirror mounted on a piezo-based translation stage (an exact replica of the ones used in FLUOR or VINCI instruments) replaces one of the flat mirrors of the delay line.
Finally, the detector used with this beam combiner is a science grade Nicmos camera with a digital gain of \SI[per-mode=symbol]{10}{\electron\per\adu}, a read-out noise of \SI{30}{\electron}, and a $45\%$ quantum efficiency.

\subsubsection{Control software}\label{SSSec:ControlSoftware}

The supervisory control software, programmed in LabView, runs on a single desktop computer.
This software interfaces with the Iki telescopes' host computers through the ethernet network, to request acquisition of new targets and to optimize the coupling into the fibres by performing raster scans of the detected flux using telescope offsets.
The software is also responsible for triggering delay line carriage moves, and re-align automatically the input fibre to output fibre coupling on both sides.
Based on an estimation of the deployed baseline vector, a sidereal target (position and velocity at a given time) is computed at \SI{1}{\hertz}, and serves as a setpoint for the control of the continuous delay translation stage operated in velocity mode.
At configuration time, the parameters of the multi-window non-destructive readout are sent, through a serial interface, to the drive electronics of the infrared camera.
Once in free-run mode, the drive electronics are responsible for generating the detector clocks, pre-conditioning, then digitizing the analog video signal.
The digital signals are sent, through a fibre optic link, to a digital acquisition card (National Instrument PCI-DIO-32HS).

Configured in a free-running mode, the detector continuously repeats a pattern of 1 reset followed by N non-destructive reads.
Each start of a pattern triggers, on an analog output card (National Instrument PCI-MIO-16E-4), a pre-computed triangular waveform driving the OPD scanner.
The result is a scanned fringe detection, whose parameters can be adjusted to match the magnitude of the observed object, and fit within the linearity regime of the detector. 
A spiral OPD pattern on top of the sidereal target, introduced at the level of the delay line (see \ref{SSSec:DelayLine}), allows searching for zero OPD until fringes appear within the scan length.
Once the various flux optimizations described earlier are completed and the fringes found, the supervisory software finally carries out the data collection sequence; taking dark frames (both telescopes offset from the target star), obtaining flux ratios from each telescope (each telescope sequentially offset from the target), and recording the fringes (both telecopes on target) as raw data in the disk archive.

\subsection{System qualification}\label{SSec:SystemQualification}

This section present the system level performance of the instrument; specifically the piston stability (Section~\ref{SSSec:InternalFringesPerformance}), and the overall sensitivity given the transmission and injection losses present (Section~\ref{SSSec:TransmissionInjectionStability}).
The model used the determine the baseline is outlined in Section~\ref{SSSec:BaselineModels}).

\subsubsection{Internal fringes performance}\label{SSSec:InternalFringesPerformance}

Before being used on sky, the J-band fibres, delay line, and beam combiner were tested independently in a Mach-Zender configuration, replacing the two telescopes by an additional J-band $2\times2$ coupler connected to the fibre pair inputs and fed with a thermal source.
This setup was not only used to measure the zero group delay position in the absence of sidereal contribution, but also to assess the OPD stability through the fibres and delay line, as installed inside CFHT.

In order to reduce the vibrations detected in the initial setup, the fibres were systematically isolated from the telescope building with tubular foam-rubber pipe insulation.
However, the measured vibration level remained high, with RMS amplitudes in the range of two waves in J-band, and frequencies of hundreds of Hertz.
As shown in Fig.~\ref{Fig:InternalFringesPsd}, different internal fringe configurations were used to better identify the origin of the measured vibrations. 
A series of measurements was taken with the CFH telescope drive and dome drive hydraulic pumps turned off.
A comparison was made between the power spectral density (PSD) of the internal fringes obtained using three configurations; a \SI{300}{\meter} fibre single-pass Mach-Zehnder, a \SI{10}{\meter} fibre single-pass Mach-Zehnder, and the delay line operated in double-pass auto-collimation.
In each case, the PSD peak occurs at \SI{\pm200}{\hertz}, which represents the fringe frequency, proper.
However, when comparing the quiet environment PSDs, the peak is disproportionately broader and lower for the double-pass auto-collimation setup compared to the single-pass configurations with longer fibre lengths.
We conclude from this that the main contributor to the interferometer instability is the delay line.
  
With the hydraulics pumps powered back on, the PSD peak was further reduced and broadened to a full width of \SI{\pm400}{\hertz}, illustrating the role of external excitation sources on the delay line vibrations.
Various modifications were then applied to the delay line, involving the installation of damping material, and the stiffening of optical elements.
This resulted in some additional performance improvement, with the PSD full width being reduced to \SI{\pm150}{\hertz} with the hydraulics pumps powered on.
 
Based on these internal performance measurements, the vibrations had a significant impact on the sensitivity of the `OHANA Iki setup.
In order to preserve the fringe contrast despite the high frequency vibrations, the measurement rate had to be kept higher than \SIrange{500}{1000}{\hertz}, resulting in a fringe peak frequency of at least \SIrange{100}{200}{\hertz}, at a scanning speed of $5$ samples per fringe.
Even though this is an acceptable limitation for a demonstration, the sensitivity of the instrument could have been improved with the addition of metrology through the fibres and delay lines and/or by significant modification of the delay line structure.

\subsubsection{Transmission and injection stability}\label{SSSec:TransmissionInjectionStability}

With the help of raster scan patterns (see Fig.~\ref{Fig:RasterScan}), the coupling at each telescope of the star light into the single-mode fibre was optimized.
Iterative adjustments of the tip/tilt, telescope focus, and focal reducer F-ratio, ensured optimal coupling of the telescope point spread function with the fibre input mode.
A similar optimization was performed between the input and output fibres of the delay line.
The stability of the injection into the single-mode fibres and the total number of photons detected at the focus of the interferometer were then measured on sky.

An observation of Antares (J=-2.73) resulted in an average detected flux of \SI[per-mode=symbol]{3500}{\electron\per\milli\second} per telescope, which corresponds to \SI[per-mode=symbol]{283}{\electron\per\milli\second} per telescope for a $J=0$ target.
Compared to the theoretical transmission error budget given in Table~\ref{tbl:TransmissionErrorBudget}, there is a measured additional loss of $\sim\times 23$.
There is no fundamental reason to incriminate the fibres: percent level transmissions have already been demonstrated in H/K band on the Keck Interferometer \citep{Perrin+2006a}.
This transmission is certainly specific to the ‘OHANA Iki experiment.
A similar transmission discrepancy of $\sim\times 20$ was also measured with the H-band fibres, confirmed by the frequency conversion experiment of \citet{Ceus+2012}, using the same H band `OHANA Iki equipment, except for the delay line and beam combiner.
Independent verifications of the delay line and fibres transmission on an internal source did not show any significant discrepancy with the transmission error budget, leaving the Iki telescopes as the most probable source of losses.
The quasi-ideal Airy pattern shown in Fig.~\ref{Fig:RasterScan} seems to rule out an image quality issue. 
The transmission of the telescopes in the near-infrared is a significant concern: designed for visible operation, they are not specified in the near-infrared by the manufacturer.
An actual measurement of this near-infrared transmission would have been better than just the characterisation of a Schmidt corrector plate supposed identical to those in the Ikis \citep[see details in][]{Baril+2010}.
Nevertheless, the impact on sensitivity from these losses did not prevent the pursuit of our interferometric objectives.

Finally, the temporal stability of the injection was verified.
We did not note any significant injection fluctuations linked to atmospheric turbulence, as would be expected from a small \SI{20}{\centi\meter} diameter telescope, when the the J-band Fried parameter is $r_{0,J}=\SI{30}{\centi\meter}$ under average \SI{1}{\arcsec} Mauna Kea seeing.
The telescopes however were impacted by wind shake when left unprotected.
Windscreens were installed to help mitigate this effect.

\subsubsection{Baseline models}\label{SSSec:BaselineModels}

Due to the lack of accurate metrology on the telescope positions, the interferometer wide angle baseline was only known to the order of a centimeter at the beginning of any observing night.
With only a pair of targets observable as a result of the limited sensitivity, a pointing model for the baseline vector could not be measured at the beginning of each night.
Therefore, an alternative method had to employed to measure the baseline.

With the internal optical path difference constant and independent of the baseline vector, the error in predicting the position of the zero optical path difference is only dependent on the contribution from the baseline error $\delta\vec{B}$ to the external delay $\vec{s}\cdot\delta\vec{B}$, for the given direction $\vec{s}$ of the observed object.
The baseline error stems from the approximate knowledge of the position of the telescopes in a newly deployed array configuration.
Once the fringes are found and tracked, the baseline error can be corrected as follows:
\begin{alignat}{2}
		                                                          \vec{s} \cdot \delta\vec{B} &= \delta\Delta,       \label{Eq:BaselinePosiitonOffset}\\
		                             \left(\vec\omega\times\vec{s}\right) \cdot \delta\vec{B} &= \frac{\partial\delta\Delta}{\partial t}, \label{Eq:BaselineVelocityOffset}\\
		\left[\vec\omega\times\left(\vec\omega\times\vec{s}\right)\right] \cdot \delta\vec{B} &= \frac{\partial^2\delta\Delta}{\partial t^2},\label{Eq:baselineAccelerationOffset}
\end{alignat}
using the value, velocity, and acceleration of the offset $\delta\Delta$, where $\vec\omega$ is the earth rotation vector.
However, since the amplitude of the earth rotation vector is very small ($\omega=\SI{7.3e-5}{\radian\per\second}$), the accuracy of the baseline error projected on the velocity vector is poor, and the projection on the acceleration is even less accurate.
For this reason, only offsets for the position and velocity projections were applied in practice.

\subsection{First contrasts measurements}\label{SSec:FirstFringes}

Given the sensitivity losses from vibrations (section \ref{SSSec:InternalFringesPerformance}), the low transmission throughput (section \ref{SSSec:TransmissionInjectionStability}), and the difficulty of finding fringes on a roughly known baseline, the demonstration proceeded using a very short baseline.
The initial installation of the two telescopes on a $\sim$\SI{1}{\meter} north-south baseline was performed outside of CFH, using crude visual sightings between the CFH building itself and the neighboring meteorological tower.
Since the telescopes could not be left unattended outside during the day, references were left on the ground in order to repeatably position them at the beginning of each observing night.
Illustrated in Fig.~\ref{Fig:FirstFringes}, the first J-band fringes were obtained on Antares ($\mathrm{J}=-2.7$), in July 2010 \citep{Dong2010}.

Following this initial demonstration, the performance of the interferometer was systematically improved by stabilizing and maximizing the coupling between the telescopes and the fibres, mitigating the impact of vibration on the fibres and the delay line, and improving the noise performance of the infrared detector.
The state of the sub-systems following this improvement campaign are described in section \ref{SSec:InstrumentDescription}.
Ultimately, the performance reached a level compatible with the observation of a few bright stars with known diameters: Antares ($\mathrm{H}=-3.5$), Arcturus ($\mathrm{H}=-2.8$), Betelgeuse ($\mathrm{H}=-3.7$), and Scheat ($\mathrm{H}=-2.1$).

In June, 2012, the H-band fringe contrast stability was assessed with observations of Antares and Scheat.
The fringe contrast was estimated using a technique similar to the integration of the fringe peak of the interferometric signal power spectral density, used for the FLUOR instrument \citep{CoudeDuForesto+1997}.
Here however, a single, instead of a triple coupler was used without real-time photometry.
Single telescope fluxes were estimated before fringe collection in order to estimate the flux ratio contrast loss.
The instantaneous photometric fluctuations were rejected by computing the normalized fringe signals $(I_1-I_2)/(I_1+I_2)$, where $I_1$ and $I_2$ are the two interferometric outputs of the 50:50 coupler.
The power spectral densities of these normalized fringe signals were then corrected for the photon noise \citep{Perrin2003} and readout noise \citep{Woillez2004} biases.
The square fringe contrast was calculated by integrating the bias-corrected mean power spectral density.

Results for the night of June 12, 2012, are shown in Fig.~\ref{Fig:FringeContrast}.
The large $\pm\sigma\approx\pm0.08$ error bars on the fringe contrast estimates result from the lack of simultaneous photometric measurements.
The estimated squared fringe contrast for Antares and Scheat are respectively 0.71 and 0.69.
Unfortunately, these observations were performed on a \SI{1}{\meter} North-South baseline; too short to detect any differential diameter.
The predicted square visibility of Antares (diameter $\SI{\sim32}{\milli\arcsec}$) and Scheat (diameter $\SI{\sim17}{\milli\arcsec}$), on the respective $\SI{\sim0.69}{\meter}$ and $\SI{\sim0.75}{\meter}$ projected baselines, are $V^2=0.995$ and $V^2=0.987$.
A $\sim\SI{4}{\meter}$ baseline would have been better suited to demonstrate visibility cross calibration.

The June 2012 run concluded the `OHANA Iki experiment, validating the technological readiness of a fibred interferometer link between the large telescopes of Mauna Kea.
Out of the three phases outlined in the original project roadmap, most of the objectives of Phases I and II have been met; adaptive optics to fibre interfacing \citep{Woillez+2003} in Phase I, and an on-sky interferometric demonstration \citep{Perrin+2003} in Phase II, albeit not using the originally planned CFH-Gemini interferometric combination.
Whereas, \citet{Perrin+2006a} successfully demonstrated on-sky, long-distance coherent transport through single-mode fibres using the Keck Interferometer, the present paper moves a step further in reporting on the successful implementation and on-sky operation of an interferometer based entirely on single-mode fibres.
These two milestones lead the way for Phase III \citep{Lai+2003}: the construction of the full `OHANA array envisioned by \citet{Mariotti+1996}.
For now, the realization of this vision remains beyond the resources of our current team supported by the observatories of Mauna Kea.
However, a fully operational `OHANA array is presently a \emph{demonstrated possibility}, for a wider community to eventually implement.

\section{An AGILIS concept, and new project?}\label{sec:Agilis}

The coherent guiding and high transmissivity properties of single-mode fibres allows a greater separation between the light collection, the light delay, and the light combination functions of an interferometer, while preserving sensitivity.
In the original `OHANA concept \citep{Mariotti+1996}, this function separation is used to easily convert, with no significant infrastructure modification, the already available telescopes on Mauna Kea into an interferometer.

Similarly, the `OHANA Iki experiment opens another perspective on this function separation: greater freedom is given to the layout of the light collection and delay functions.
In the previous section, this capability was partly illustrated when the Iki telescopes had to be setup outside for an observing night: any kind of baseline could have been chosen, without requiring any baseline-specific change in the delay lines setup.
In this section, we explore this idea further under the concept name ``Agile Guided Interferometer for Long-baseline Imaging Synthesis'' or AGILIS.

\subsection{Agile (u, v) coverage}\label{SSec:AgileUVCoverage}

The main first-order requirement for an imaging optical interferometer is a dense (u,~v) plane coverage, to be achieved in a minimum amount of time for evolving targets or for efficiency reasons \citep[see e.g.][]{Berger+2012}.
Measurements of the complex visibility from individual observations can be recombined \textit{a posteriori} in an image reconstruction process similar to that used in radio-interferometry \citep[e.g.][]{Sutton&Wandelt2006}.

An imaging optical interferometer must be cophased.
To achieve a good limiting magnitude, the fringe sensor employed to cophase the array must operate on the shortest baselines exclusively, where a high fringe contrast and a reasonable flux per baseline are maintained.
As a reminder, and considering a uniform disk object as an example, the maximum $V^2$ in the second lobe is on the order of $\sim 0.17$, corresponding to a sensitivity loss of $4.4$ magnitudes, under the assumption that photon noise still dominates.
The phasing of the longer baselines can only be achieved, without a significant sensitivity impact, by bootstrapping small baselines where the object is unresolved \citep{Roddier1988, Armstrong+1998, Hajian+1998}.
In this configuration, the shortest baselines should match the overall size of the observed object, whereas the longest baselines should be adapted to the size of the smallest details desired in the reconstructed image. 
Meeting all these requirements with a fixed telescope array, for a wide range of objects, is not possible.
For this reason, optical interferometers typically use relocatable telescopes to adapt the measured spatial frequencies to the size of the observed object, and sequentially pave the (u,~v) plane, as at IOTA, VLTI, or the planned MROI \citep[resp.][]{Carleton+1994, vonDerLuehe+1997, Buscher+2013}.

With the proposed AGILIS flexibility, the use of single-mode fibres allows rich aperture synthesis with telescopes that can be moved to any location on a terrain, satisfying the requirements above.
A global scaling of the array can match the size of the object; a relative scaling between north and east directions can compensate the projection effects of target declination versus instrument latitude and hour angle.
Therefore, the array can be optimized both in sensitivity with its shortest baselines, and in angular resolution with its longest baselines.

Innovative array geometries can be considered without the cost impact of a beam routing infrastructure in bulk optics.
The regular Y configuration \citep{Buscher+2000}, where the stations are aligned in each of the three arms in order to simplify beam routing, can now be modified to remove the array redundancy, as done in \citet{Ireland+2014}, increasing the instantaneous (u,~v) coverage by $+25\percent$ for a 10 telescope configuration (see Fig.~\ref{Fig:ArrayConfigurations}).
The non-redundant circular arrays \citep{Labeyrie+1986, Labeyrie1998} become practical, with fibre links running from the individual telescopes to a central delay and beam combination location.
Called hypersynthesis in \citet{Vivekanand+1988}, continuous (u,~v) coverage adaptation, with telescopes moving while observing, could even be considered.

The benefit gained in freeing the telescope layout is most obvious when combining a small number of telescopes.
The array can be configured for a dense two-dimensional (u,~v) coverage, or it can be dedicated to reaching the longest baseline possible with a regularly spaced linear array (see Fig.~\ref{Fig:ArrayConfigurations}).
This amounts to a trade-off between dense (u,~v) coverage for instantaneous low angular resolution with high image quality, and sparse (u,~v) coverage for sequential high angular resolution but poor image quality.

\subsection{Delay compensation}

All the fibres connecting the telescopes to the delay lines must have the same length, in order to balance their inherent chromatic dispersion \citep{Vergnole+2004,Vergnole+2005,Kotani+2005,Anderson+2014}.
Doing so, all the telescopes of the array are at the same constant optical path distance from the delay lines.
This is an improvement over bulk optic implementations, where the delay lines are often used to compensate for the optical distance of the telescope station, reducing the stroke available for pointing the interferometer.
In the fibred implementation, the delay lines do not need to provide any differential OPD for an horizontal array and a target at Zenith.
Their stroke is fully dedicated to the compensation of the external delay increasing with the zenith angle.
For a \SI{340}{\meter} baseline, and a zenith angle of \SI{60}{\degree}, a \SI{147}{\meter}-long double-pass delay line is sufficient.
In the case of a Y aligned configuration (see Fig.~\ref{Fig:ArrayConfigurations}), with the same maximum baseline of \SI{340}{\meter} as for the future MROI array, the delay lines have to be \SI{97}{\meter} longer physically, when the beams are directly routed from the telescopes, to the delay lines, through the central point of the array.
Installing fibres on such an array would reduce by $47\percent$ the delay line stroke requirement.

Requirements on the delay line stroke can still be reduced further.
As already noted by \citet{Vivekanand+1988} for a fibred array, no optical path delay is needed for telescopes contained in a plane orthogonal to the line of sight. 
The topography of the terrain combined with the mobility of the telescopes can be used to reduce or eliminate the optical path delay needed to point the array.
Movable telescopes could be set on a depression, as shown in Fig.~\ref{Fig:AgilisTopography}, where the source direction is locally perpendicular to the surface.
Then, delay lines are only required to compensate for Earth rotation during an exposure.
The maximum stroke of the delay lines therefore gives the maximum time during which a configuration can be used before telescopes are relocated to another optimum location.
As an illustration, for a kilometric east-west baseline at latitude zero observing an object at zenith (i.e. the worst yet realistic geometric configuration for Eq.~\ref{Eq:BaselineVelocityOffset}), the fringe velocity is \SI{7.3}{\centi\meter\per\second}, which corresponds to an autonomy of at least \SI{10}{\minute} for a \SI{50}{\meter} optical path delay line.

The idea of using terrain to support an interferometer or dish is not new.
It has been demonstrated by the Arecibo radio telescope and has been proposed by \citet{Labeyrie&Mourard1990} for the Lunar Optical Very Large Interferometer.
More recently  \citet{Labeyrie+2012_Ubaye} have used the topography of the Vall\'{e}e de l'Ubaye to install an equivalent to the CARLINA hypertelescope, previously demonstrated by \citet{LeCoroller+2012}.
However, in all these cases, only a fraction of the primary spherical mirror can be used in a given pointing direction.
For the hypertelescope, this is set by the amount of spherical aberration that the Mertz corrector can tolerate.
For an AGILIS concept using the terrain topography and movable telescopes, all telescopes are active for a given observation and short delay lines play the role of the Mertz corrector of a hypertelescope.

\subsection{Modal transport}

The coupling loss between the airy pattern at a telescope focus, and the fibre fundamental mode \citep{CoudeDuForesto+2000} can be overcome.
An off-axis telescope, without central obscuration, can already increase the coupling efficiency to $78\percent$.
Following a technique proposed by \citet{Guyon2003}, the entrance pupil of the individual telescopes can even be losslessly modified to match the mode profile of the fibres \citep{Jovanovic+2016}.
A coupling of $100\percent$ can virtually be reached, limited only by the size of the mode matching optics.
This mode matching approach removes the small low-flux phase estimation disadvantage of single-mode over multi-mode recombination identified in \citet{Tatulli+2010}. 
Once the light is injected into the fibre at the telescope focus, the fundamental mode can virtually be preserved all the way to the cophaser and science instrument.
This preservation includes the polarization state, either using silicate polarisation maintaining fibres for J and H bands, or low bi-refringence fluoride glass fibres with passive polarisation control \citep{Lefevre1980} for K band.

Ideally, the delay function would be performed with the very same fibres.
However, the intrinsic dispersion of those currently available \citep{Kotani+2005,Vergnole+2004,Vergnole+2005} prevents the generation of large optical path delays by stretching or adding/removing fibre sections, that would be usable over the typical bandwidth of optical atmospheric windows.
Only broad band delays on the order of a few centimeters have been demonstrated in laboratory experiments \citep{Huss+2001}, or are being integrated in upcoming instruments \citep{Eisenhauer+2011}.

The current alternative is to use free-space delay lines.
The fundamental mode of a step-index single-mode fibre is close to having a Gaussian profile \citep[][]{Gloge1971}, which is an invariant of diffraction.
The only effect, the curvature resulting from the propagation, can easily be compensated with a defocus of the collimating interfaces with the fibres.
A double-pass delay line coupled with single-mode fibres can therefore be designed with virtually no diffraction losses.

Since the objective is to combine a large number of telescopes, the footprint of a transverse section of the delay line must be kept to a minimum.
This is achieved with the Gaussian beam waist located at mid-range between the fixed collimation/injection optics and the mobile delay-line cart.
This configuration requires the capability to track the waist and cart locations with mobile optics at the level of the collimation, injection, and cart.
Such variable pupil relay optics are already in operation in the delay line carts of VLTI \citep{Ferrari1998}.
The implementation could however be simplified by locating the waist at the level of the cart itself, requiring tracking optics only at the level of the fixed collimation and injection, leaving the delay cart optics completely passive.

A kilometric baseline at a zenith angle of \SI{45}{\degree} requires a delay of \SI{700}{\meter}, corresponding to a maximal physical length of \SI{350}{\meter} in double-pass.
In order to keep the Rayleigh distance $z_R=\pi \omega_0^2/\lambda$ larger than this physical length, the waist radius $\omega_0$ of the Gaussian beam at the delay line cart must be larger than \SI{1.57}{\centi\meter} in K band ($\lambda=\SI{2.2}{\micro\meter}$), or \SI{0.82}{\centi\meter} in R band ($\lambda=\SI{0.6}{\micro\meter}$).
For a transmission of $99\percent$ of the beam power, taking into account the beam divergence after one Rayleigh distance propagation, the optics diameter must be larger than \SI{6.75}{\centi\meter} in K band, or \SI{3.5}{\centi\meter} in R band.
For an estimation of the required volume, each delay line would fit in a cylinder of \SI{15}{\centi\meter} diameter, the double of the optics diameter including margins, so that one hundred of such delay lines could be hexagonally packed, under vacuum, in a cylinder of \SI{1.6}{\meter} diameter.

Finally, a spin-off from the field of femto-second lasers, ultra highly reflective broad-band dielectric mirrors with low group delay dispersion are becoming available \citep{Trubetskov+2010}.
With these mirrors, fibre-pigtailed, multiple-pass commuted delay lines \citep{Ridgway+2003}, could significantly reduce the physical length of the delay line, at the cost of an increased cross-section.
By allocating to a given beam only the exact volume required to produce the necessary delay, a bank of commuted delay lines could have an overall volume half the size of a traditional bank of double-pass delay lines.

\subsection{Co-phasing the array}\label{SSec:CophasingArray}

As already mentioned in section \ref{SSec:AgileUVCoverage}, in order to preserve sensitivity, the co-phasing of the array can only happen on the shorter baselines where the object is unresolved and the fringe contrast high.
On the contrary, the imager has to measure all the baselines as corrected by the short baselines' co-phaser.
In this section, we investigate how the performance of the co-phaser on these shorter baselines affects the performance of the imager.

The co-phasing loop is modelled as illustrated in Fig.~\ref{Fig:ControlLoop}, where only piston perturbations and photon noise are considered.
This simplification is not far from reality; recent progress with near-infrared detectors has made photon counting available \citep{Finger+2014}.
It is intuitive to consider a fringe sensor operating in the photon noise regime in order to achieve reasonable performance on the longest imaging baselines.
The fringe sensor converts the atmospheric piston residual error $e$ coming from the individual telescopes into differential piston with noise $n$ on each of the short fringe sensing baselines.
$\mathrm{T2B}$ represents the interaction matrix of the telescope piston phases projected onto short baselines, and depends only on the geometry of the baselines established between the different telescopes.
Because all the baselines are measured independently, the reconstructor $\mathrm{B2T}$, pseudo-inverse of $\mathrm{T2B}$, converts the differential pistons back into the individual pistons, with the exception of the global piston.
The controller $S$ integrates the reconstructed piston error and generates the piston correction $c$ associated with each telescope.
To simplify the control diagram, the time delays introduced by sensors and actuators are included in the controller model $S$.
The imager is modeled with the $\mathrm{T2I}$ matrix that generates, from the telescope residual piston error $e$, the differential piston $i$ on all the imaging baselines.

Based on this model, the residual piston error $e$ can be derived from the atmospheric disturbance $d$ and the noise $n$ as follows:
\begin{equation}
	e \sim \left(\frac{1}{1+S}\right)d - \left(\frac{S}{1+S}\right) \mathrm{B2T} \times n.
\end{equation}
This expression was obtained by observing that $\mathrm{T2B}\times\mathrm{B2T}$ is the identity, provided that the piston mode is discarded as not observable.

From the above, it becomes clear that the geometry of the fringe sensor does not directly influence the rejection of the atmospheric disturbance.
However, it plays a role in the way noise affects the control loop.
In order to study the noise propagation further, a few fringe sensing configurations were simulated: Y aligned, circular, and hexagonal.
Following the bootstrapping requirement of section \ref{SSec:AgileUVCoverage}, each configuration is built around the same unit fringe sensing baseline length, where the observed object is unresolved.
Consequently, the noise on the differential piston is strictly related to the flux contributing to a given baseline, which can be derived from the fringe tracking geometry and the assumption that all telescopes provide the same amount of light.
Considering the flux imbalances to be insignificant, the associated contrast losses are also neglected.
In this case, the phase errors $\delta\phi_I$ on the imaging baselines are related to the phase errors $\delta\phi_B$ on the fringe tracking ones by:
\begin{equation}
	\delta\phi_I = \mathrm{T2I} \cdot \mathrm{B2T} \cdot \delta\phi_B 
\end{equation}
Since the fringe tracking phase errors are uncorrelated and inversely proportional to the number of photons $N_b$ measured on the baseline $b$, the variance of the phase errors $\sigma^2_i$ on a baseline $i$ is expressed as:
\begin{equation}
	\sigma^2_i = \sum_b\;(\mathrm{T2I} \cdot \mathrm{B2T})^2_{ib} 1/N_b
\label{Eq:BaselineNoise}
\end{equation}
where $N_b$ is the number of photons contributing to the phase measurement on baseline $b$.

The results of these simulations are shown in Fig.~\ref{Fig:NoisePropagation}, representing the differential piston noise on the imaging baselines.
As one might expect, the longest baselines are the most affected because they rely more on bootstrapping.
The geometry of the array also has an influence on the noise propagation.
The results shown in Fig.~\ref{Fig:NoisePropagation} can be explained by how the fringe tracker connects the telescopes together.
The Y and circular geometries are mainly constructed with each telescope connected to two other telescopes.
These arrays are effectively one dimensional.
The hexagonal array, where each telescope is connected to three other telescopes, is two-dimensional.
The longest baselines are the result of many different bootstrapping paths, whereas there is only one for the Y array.
Note that for the circular array there are two paths, clockwise and counter-clockwise, to connect the longest baselines, therefore reducing the noise propagation as shown in Fig.~\ref{Fig:NoisePropagation}. 

In section \ref{SSec:LimitingMagnitude}, the telescopes were adapted to the seeing strength, so as to deliver a Strehl compatible with fringe tracking.
It is however interesting to consider the impact of a flux dropout for one of the telescopes within the array.  One might expect that the consequences are much less drastic for two-dimensional compared with one-dimensional arrays.
The loss of a telescope on a Y array breaks the array into two independently phased groups of telescopes (three if the dropped telescope is the central one).
With the slight redundancy of the clockwise and counter-clockwise paths, the circular array fares better.
The array stays phased but the noise propagation around the dropped telescope increases because the phasing now involves the full chain around the missing telescope.
As suggested by \citet{Petrov+2016}, redundancy can be added to one-dimensional arrays by additional fringe tracking between the second nearest telescopes.  
To preserve the sensitivity, the longest fringe tracking baseline length cannot increase (see section \ref{SSec:AgileUVCoverage}).
Therefore, the minimum distance between telescopes needs to be shortened by a factor two, increasing the overall telescope density, but without benefiting the image reconstruction.
Rather than increasing the telescope density, the flux dropout issue would be more efficiently resolved by increasing the adaptive optics correction order at the telescopes.

\subsection{Limiting magnitude}\label{SSec:LimitingMagnitude}

As mentioned in section \ref{SSec:AgileUVCoverage}, the cophaser sets the limiting magnitude of the imaging interferometer, which is related to the amount of coherent flux collected at each telescope.
Since, larger telescope diameters make $(u,v)$ coverage agility harder to achieve, only smaller telescopes with correction of the atmospheric tip/tilt are considered in the following sensitivity analysis.

The strength and speed of the atmospheric seeing drive the sensitivity of the interferometer.
At Cerro Paranal, \SI{5}{\meter} above ground, seeing strength and coherence time have been measured \citep{Sarazin+2008} to \SI{1.0}{\arcsec} and \SI{2.5}{\milli\second}, at a wavelength of \SI{500}{\nano\meter}, in median conditions.
With telescopes most probably located closer to the ground, and therefore impacted more strongly by ground layer turbulence, we consider less favorable seeing conditions at \SI{1.5}{\arcsec} and \SI{1.5}{\milli\second}.

The maximum single-mode fibre coupling efficiency for a tip/tilt corrected telescope is reached around $D=4r_0$ \citep{Shaklan&Roddier1988}.
However, the limiting magnitude of a co-phased array is set by the capability to continuously estimate the atmospheric piston.
For this, a minimum instantaneous Strehl is required.
As shown in Fig.~6 of \citet{Tatulli+2010}, $D=4r_0$ with tip/tilt correction leads to a mean Strehl (or coupling efficiency) of $|\rho_i|\sim25\%$, and therefore significant flux dropouts.
Instead, at $D=2r_0$ with tip/tilt correction, the mean Strehl increases to $|\rho_i|\sim65\%$, and the flux dropouts are significantly reduced, which improves the sensitivity of the array, relative to the telescope size\footnote{In presence of Kolmogorov turbulence, the mean phase variance after tip/tilt correction is $\Delta_3=0.134(D/r_0)^{5/3}$ \citep{Noll1976}, corresponding to a Strehl $\rho=e^{-\Delta_3}$ of $25\percent$ and $65\percent$, at respectively $D/r_0=4$ and $D/r_0=2$.}.
This is the configuration we assume.
Following \citet{Colavita+1999}, the exposure time of the fringe sensor is set to the atmosphere coherence time $\tau_0$, in order to continuously follow the atmospheric piston without introducing unwrapping errors.
A minimum required signal to noise ratio of $\mathrm{SNR}_\phi=3$ on the phase estimation (phase variance $\sigma_\phi^2 = 1/\mathrm{SNR}_\phi^2 = \SI{0.11}{\radian\squared}$) sets the number of coherent photons ($NV^2$) and therefore the limiting magnitude.
For each of the J, H, K, and R operating bands, Table~\ref{Tbl:Sensitivity} converts the atmospheric turbulence (Fried diameter $r_0$ and coherent time $\tau_0$) into the interferometer parameters (respectively, diameter $D$ and fringe sensor frequency $f$), which determine the limiting magnitude for a realistic level of transmission assumed at $5\%$. 

\begin{table*}[ht]
\centering
\begin{tabular}{cc|cccc|c}
\hline
                              &              &     K     &     H     &      J    &      R    & Unit \\
\hline
\hline
Wavelength                    &  $\lambda$   &   2.22    &    1.60   &    1.26   &    0.64   & \si{\micro\meter} \\
Bandwidth        & $\Delta\lambda/\lambda$   &   0.23    &    0.23   &    0.16   &    0.23   &  \\
Zero magnitude flux           &    $m_0$     &    670    &    1080   &    1600   &    3080   & \si{\Jansky} \\
\hline
Fried parameter               & $r_0$        &     41    &     28    &     22    &      7    & \si{cm} \\
Coherence time                & $\tau_0$     &    8.9    &    6.1    &    4.7    &     1.5   & \si{ms} \\
\hline
Telescope diameter ($2r_0$)   &  $D$         &     81    &     56    &     43    &     14    & \si{cm} \\
Fringe sensor frequency ($1/\tau_0$) & $f$   &  112.7    &  165.1    &  211.8    &   666.7   & \si{Hz} \\
\hline
Photons per exposure          & $N_{h\nu}$   &  10.7\E{6} &  5.5\E{6} &  2.7\E{6} & 0.24\E{6} & \si{\photon} \\
\hline
\hline
Limiting magnitude            & $m_{100\%}$  &   13.3    &    12.6   &   11.8    &    9.2    & mag \\
     (phase SNR of 10)        & $m_{  5\%}$  &   10.1    &     9.4   &    8.6    &    5.9    & mag \\
\hline
\end{tabular}
\caption{Estimation of the limiting magnitude $m_T$ in each of the K, H, J, and R bands.
The assumptions are a seeing strength of \SI{1.5}{\arcsec}, a seeing speed of $\tau_0=\SI{1.5}{\milli\second}$.
From these, the telescope diameter $D$ and the fringe tracker rate $f_{FT}$ can be deduced at each wavelength.
The limiting magnitude $m_T$ is given in each band considering a reasonable $T = 5\%$ total transmission to the fringe tracker ($T = 100\%$ gives an indication of the transmission loss effect).}
\label{Tbl:Sensitivity}
\end{table*}

\section{Conclusion}\label{Sec:Conclusion}

The `Ohana Iki project, initially a pathfinder experiment for the development of the `Ohana concept on Mauna Kea \citep{Mariotti+1996}, exemplified the convenience of using singe mode fibres in deploying flexible interferometric arrays.
A description of this two-telescope experiment, a proof of technical feasibility, was followed by an outline of the key elements necessary to extend this idea to a larger array of telescopes: the AGILIS concept.
The use of single mode fibres to link the light collection, delay, and combining functions of the interferometer illustrated the flexibility of configuring the array, while preserving sensitivity.
We deliberately did not settle on one technological approach for all aspects of the interferometer in order to illustrate the concept's flexibility.

We also envisage AGILIS as an array that is intended to grow progressively with time.
Starting with a small number of telescopes, it begins by capitalizing on the mobility of the telescopes to efficiently pave the UV plane.
As telescopes are added, the imaging capabilities would benefit from a phased array that is non-redundant, similar to the FIRST concept \citep{Perrin+2006b,Lacour+2007}.
With further expansion, the capability to take snapshot images of complex objects will come within reach.
In that case, the imager should be based on a homothetic remapping or a densification of the entrance pupil to directly produce an image at the focal plane.

Lastly, AGILIS would be an excellent platform for technological demonstrations in the context of the future Planet Formation Imager \citep{Monnier+2014,Kraus+2014,Ireland+2014}, or a possible way of converting the light beamer of breakthrough starshot \citep{Lubin2016} into a stellar interferometer.

\begin{acknowledgements}
Much of the work on the `OHANA Iki demonstration was carried out by students from École Polytechnique, and the authors wish to thank Benjamin Lenoir, Nicolas Clerc and Aurélien Bocquet for their contribution to this project, and Flora Bouchacourt and George Zahariade for their early contribution to the two `Ohana Iki telescopes.
The J and H band silicate fibres were integrated and characterised at XLIM with the help of Laurent Delage and Sébastien Vergnole.
Installed in the Coudé room of the CFH Telescope, the `Ohana delay line was built by the Technical Division of the Institut National des Science de l'Univers (INSU).
The `Ohana Iki demonstration would never have been possible without the unfaltering support of the Canada-France-Hawaii and W. M. Keck Telescopes.
We finally thank Poli'hau for the snow she never failed to gift us.
\end{acknowledgements}

\begin{figure}[h]
\includegraphics[width=\linewidth]{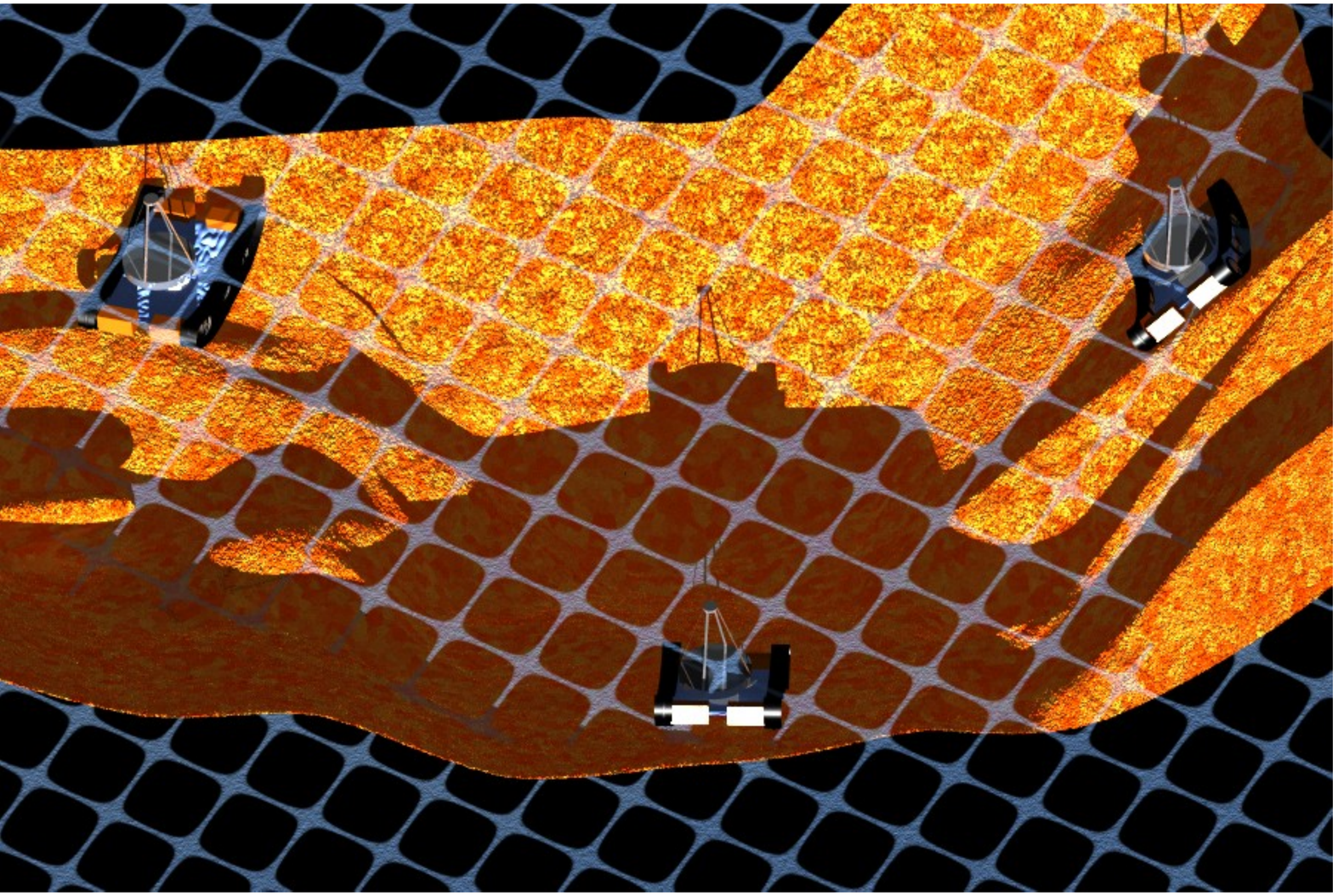}
Artist conception of a possible implementation of the early phase of AGILIS, in which a few autonomous (and self-propelled) telescopes could be used to reconfigure themselves to sequentially fill the u-v plane, using the concave (or convex) nature of the terrain to minimize the delay line requirements.
\label{Fig:AgilisVsGalileo}
\end{figure}

\bibliographystyle{aa}
\bibliography{AGILIS}

\onecolumn

\begin{figure}
\centering
\includegraphics[width=\linewidth]{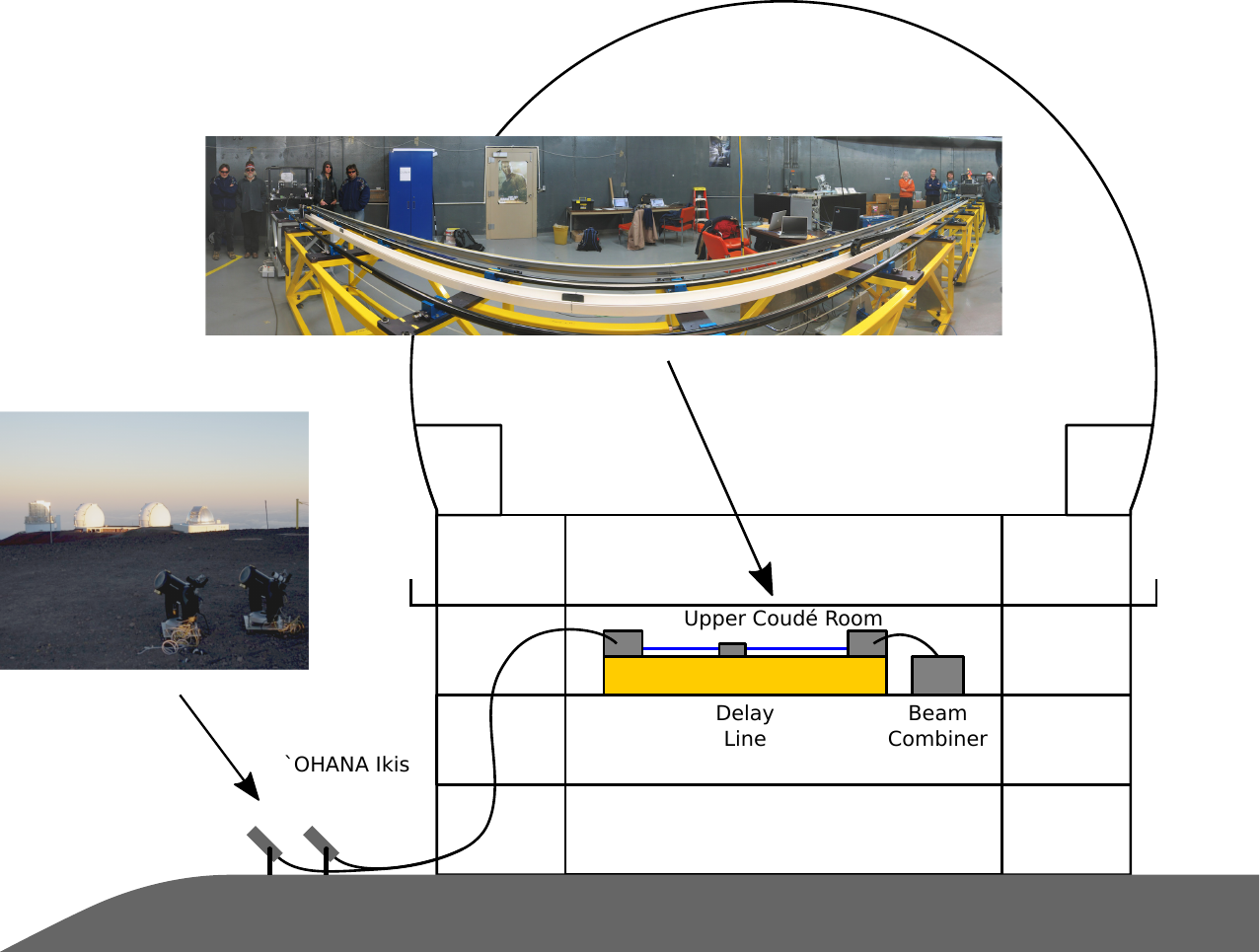}
\caption{Layout and pictures of the `OHANA Iki demonstration.}
\label{Fig:OhanaIkiLayout}
\end{figure}



\begin{figure}
\centering
\includegraphics[width=\linewidth]{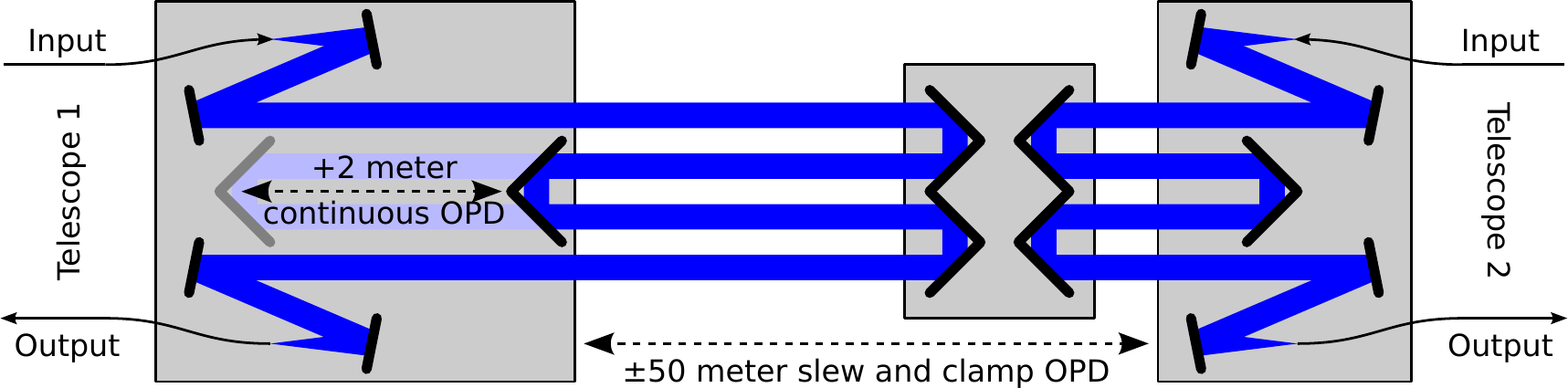}
\caption{`OHANA Delay Line. The delay line is made of two optical tables, joined by \SI{12.5}{\meter} long rails, supporting a mobile central carriage. The central carriage, used in a $\times 8$ optical pass setup, provides a slew and clamp adjustable optical path difference of \SI{\pm50}{\meter}. A \SI{1}{\meter} long translation stage on one of the optical table, provides a continuously adjustable \SI{2}{\meter} OPD.}
\label{Fig:DelayLine}
\end{figure}



\begin{figure}
\centering
\includegraphics[width=\linewidth]{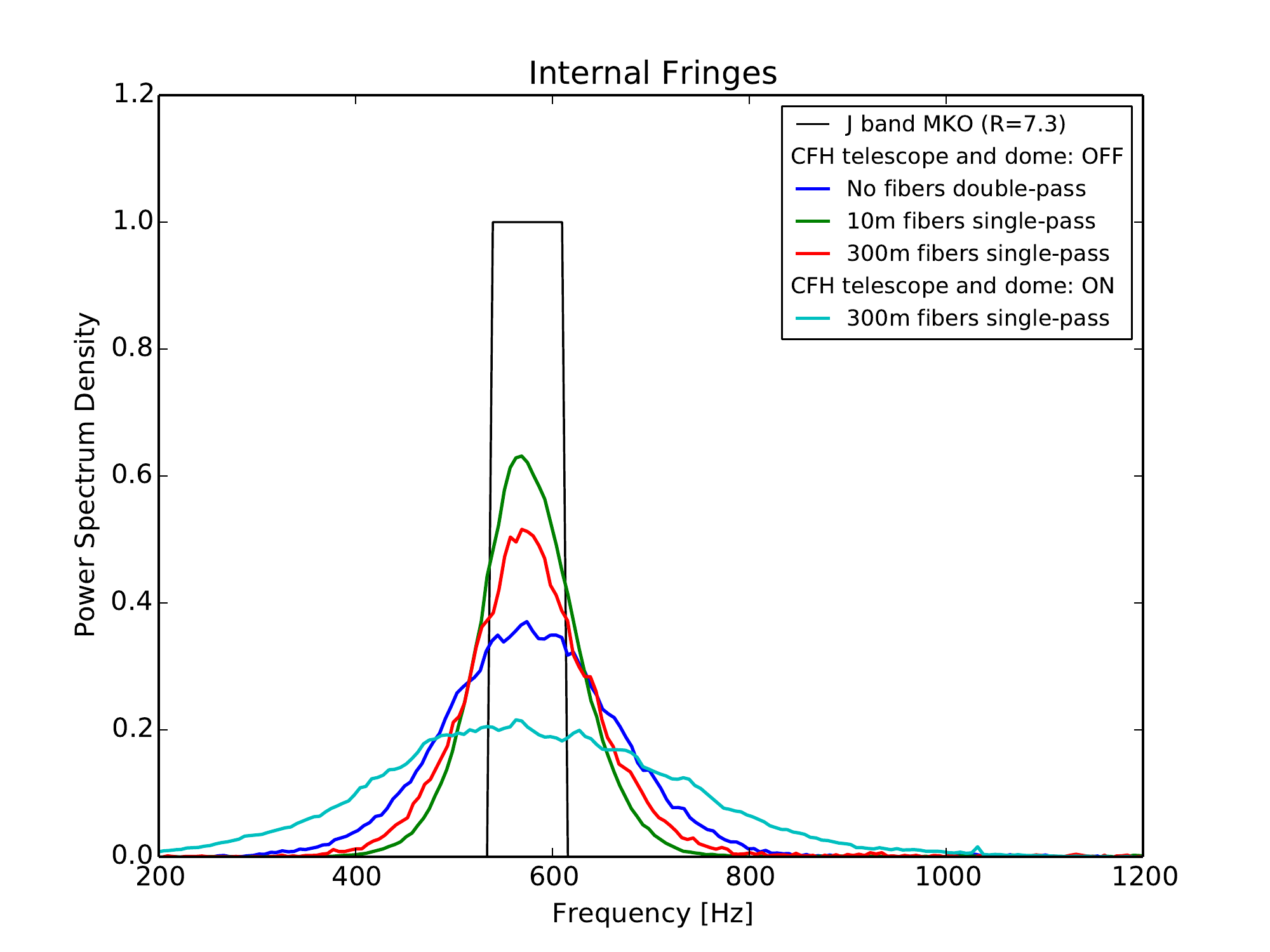}
\caption{Power spectral density (PSD) of internal fringe signals for various configurations, compared to an ideal J band filter (black): double-pass Michelson with delay-line only (blue), single-pass Mach Zehnder with 10m fibres and delay line (green), and single-pass Mach Zehnder with 300m fibres and delay line (red). The fringe peak is significantly broadened by the vibrations of the delay line; the additional vibrations in the \SI{10}{\meter}-long and the \SI{300}{\meter}-long fibres are comparably smaller.}
\label{Fig:InternalFringesPsd}
\end{figure}

\begin{figure}
\centering
\includegraphics[width=\linewidth]{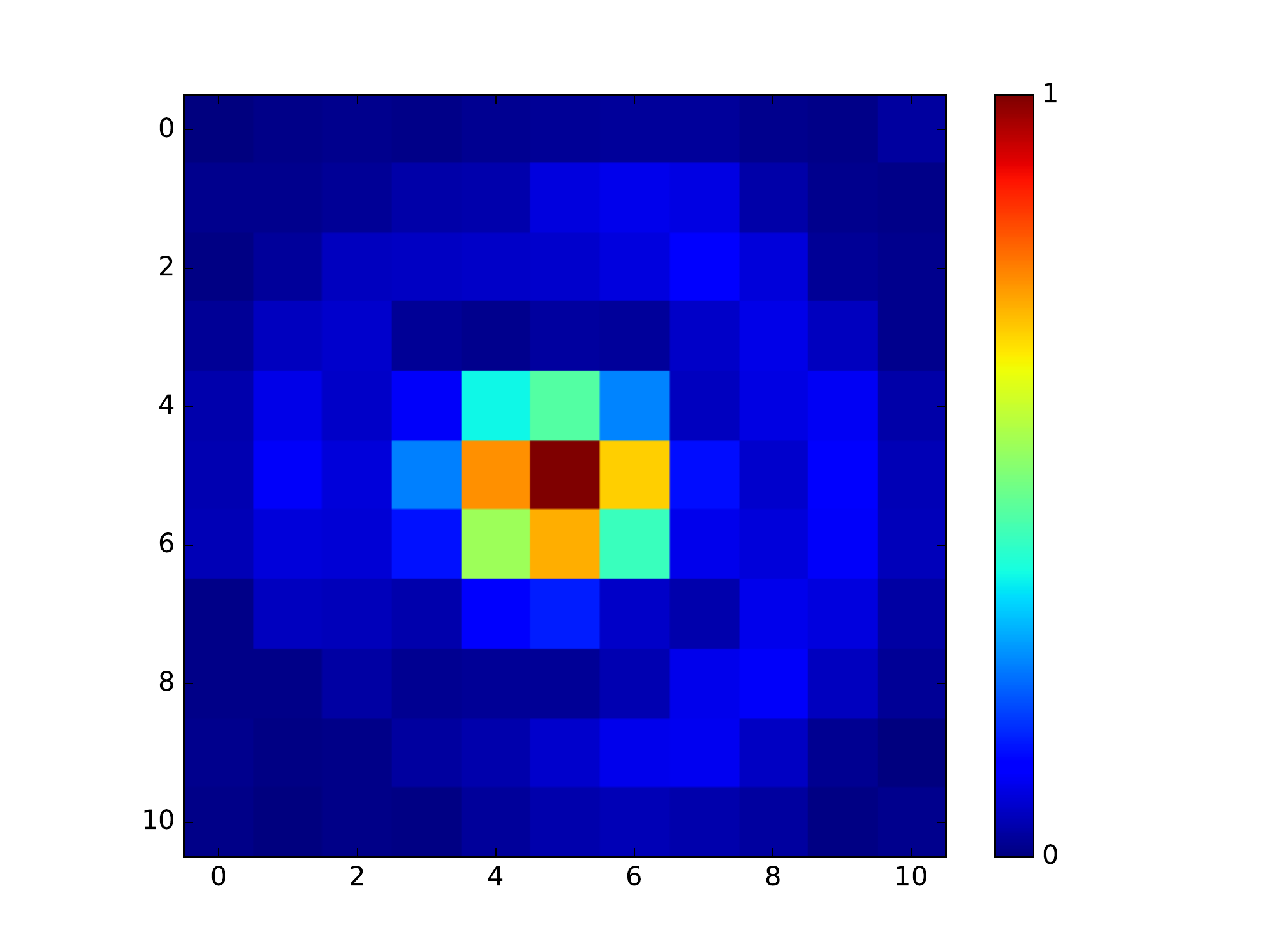}
\caption{Typical raster scan image obtained by exploring the field with telescope offsets, and recording the flux detected at each field position. A prominent first Airy ring, resulting from the significant central obscuration of the Iki telescopes, is visible around the central core.}
\label{Fig:RasterScan}
\end{figure}

\begin{figure}
\centering
\includegraphics[width=\linewidth]{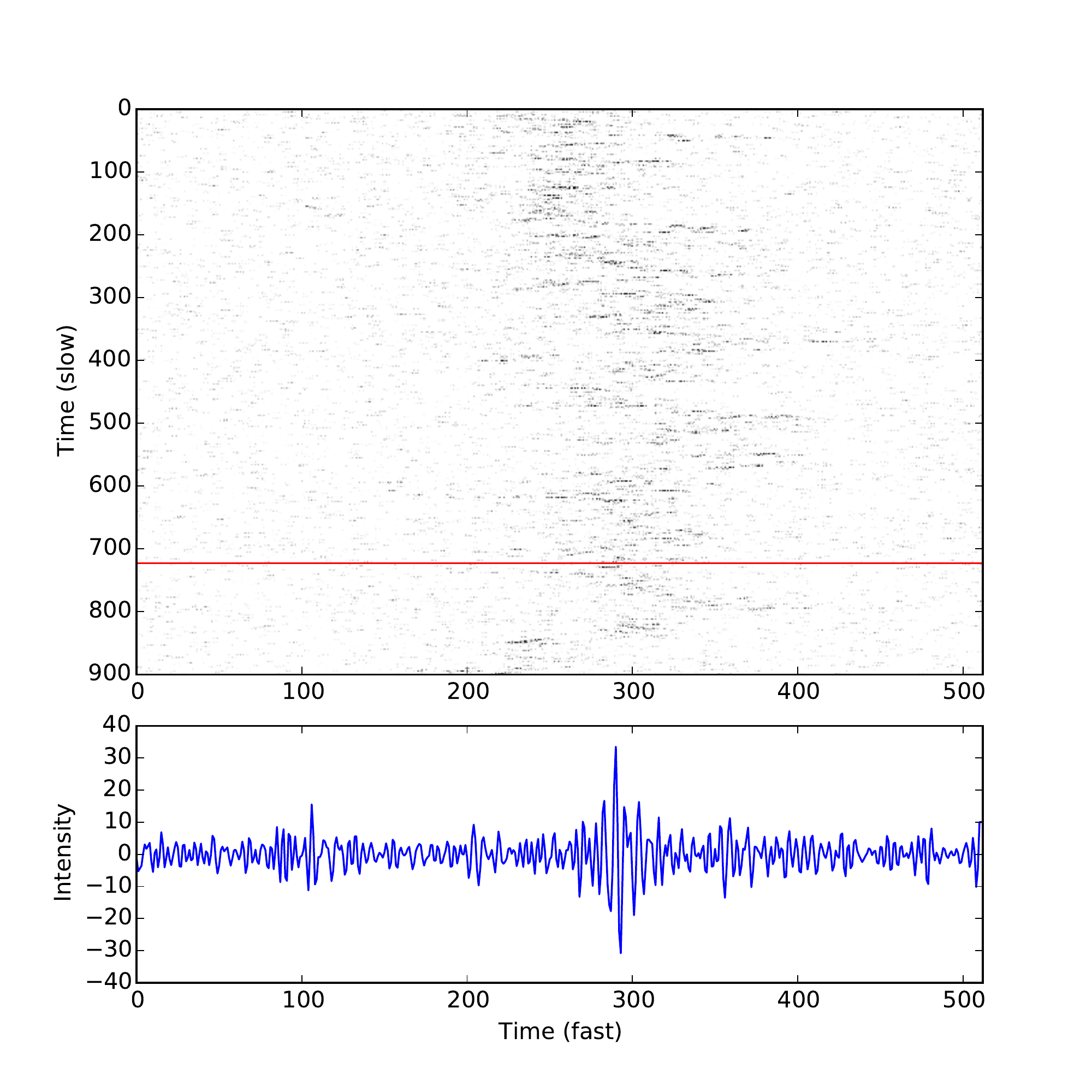}
\caption{First J-band `OHANA Iki fringes. The temporal drift of the zero optical path difference was corrected to keep the fringe packet close to the center of the scan. \textbf{Top}: Waterfall intensity display. \textbf{Bottom}: Single fringe scan, corresponding to the horizontal line marked on the waterfall display above.}
\label{Fig:FirstFringes}
\end{figure}

\begin{figure}
\centering
\includegraphics[width=\linewidth]{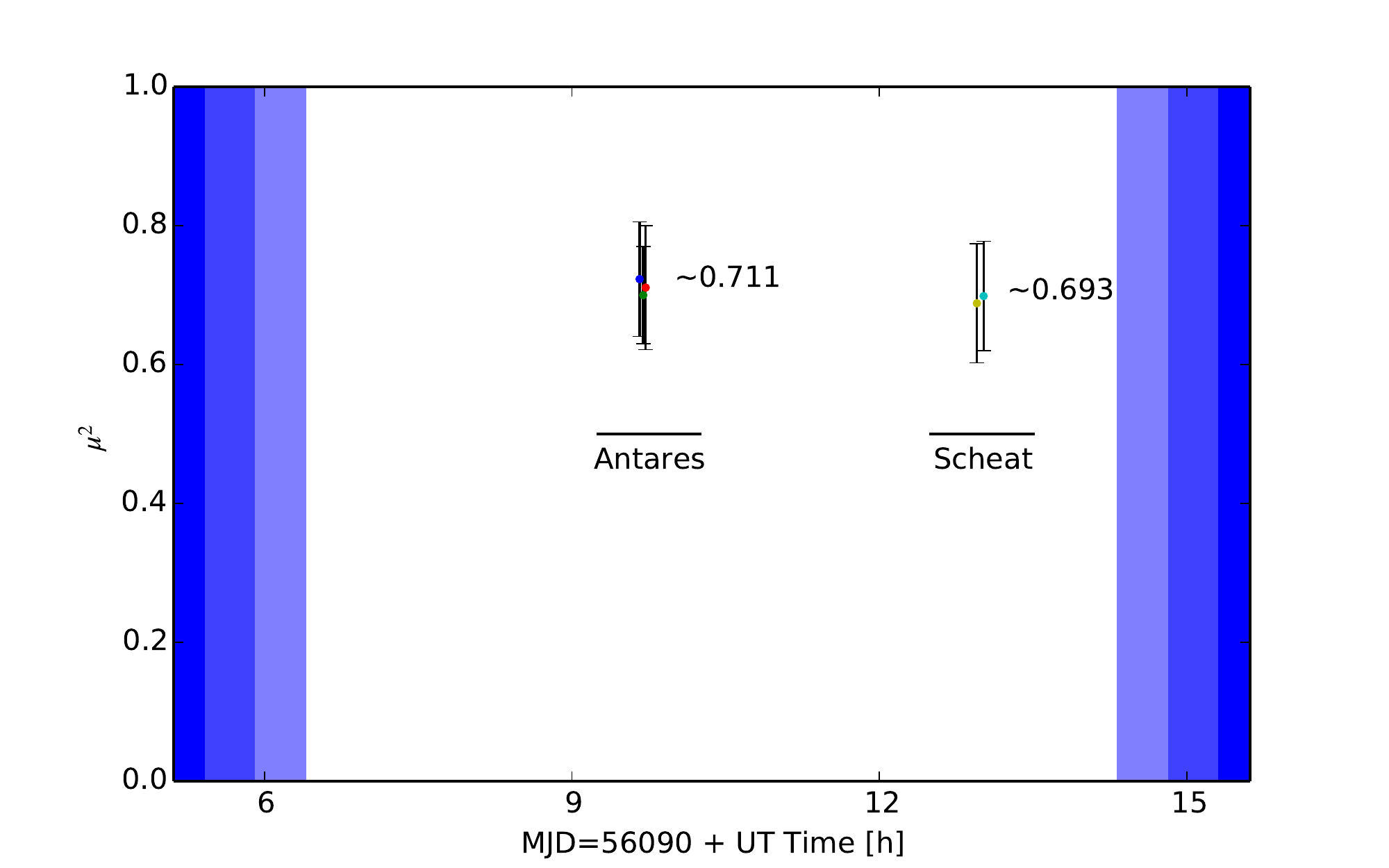}
\caption{H-band fringe contrasts $\mu^2$, and $\pm\sigma$ error bars, measured on Antares and Scheat, for the night of June 12, 2012 (MJD=56090).}
\label{Fig:FringeContrast}
\end{figure}

\begin{figure}
\centering
\includegraphics[width=\linewidth]{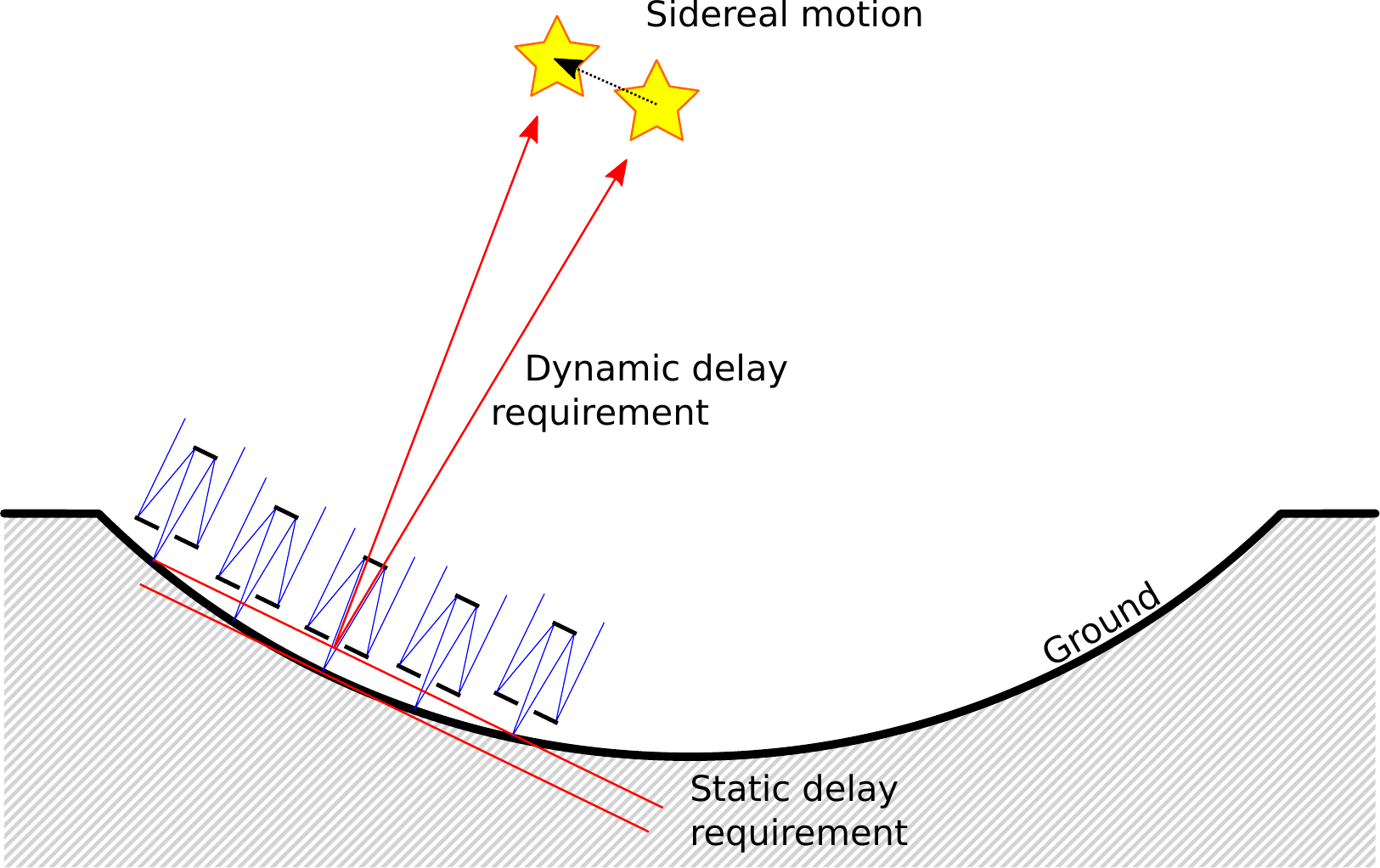}
\caption{An AGILIS concept using the terrain to minimize the delay line stroke requirement. The interferometer is set in a spherical-like depression. Telescopes are moved to a portion of the depression where the source direction is perpendicular to the surface. The requirement on the delay lines stroke is set by: 1) a static term associated to the departure of the local ground surface from a plane, and 2) a dynamic term associated to the desired pointing autonomy, derived from the observed target sidereal motion.}
\label{Fig:AgilisTopography}
\end{figure}

\begin{figure}
\centering
\includegraphics[width=\linewidth]{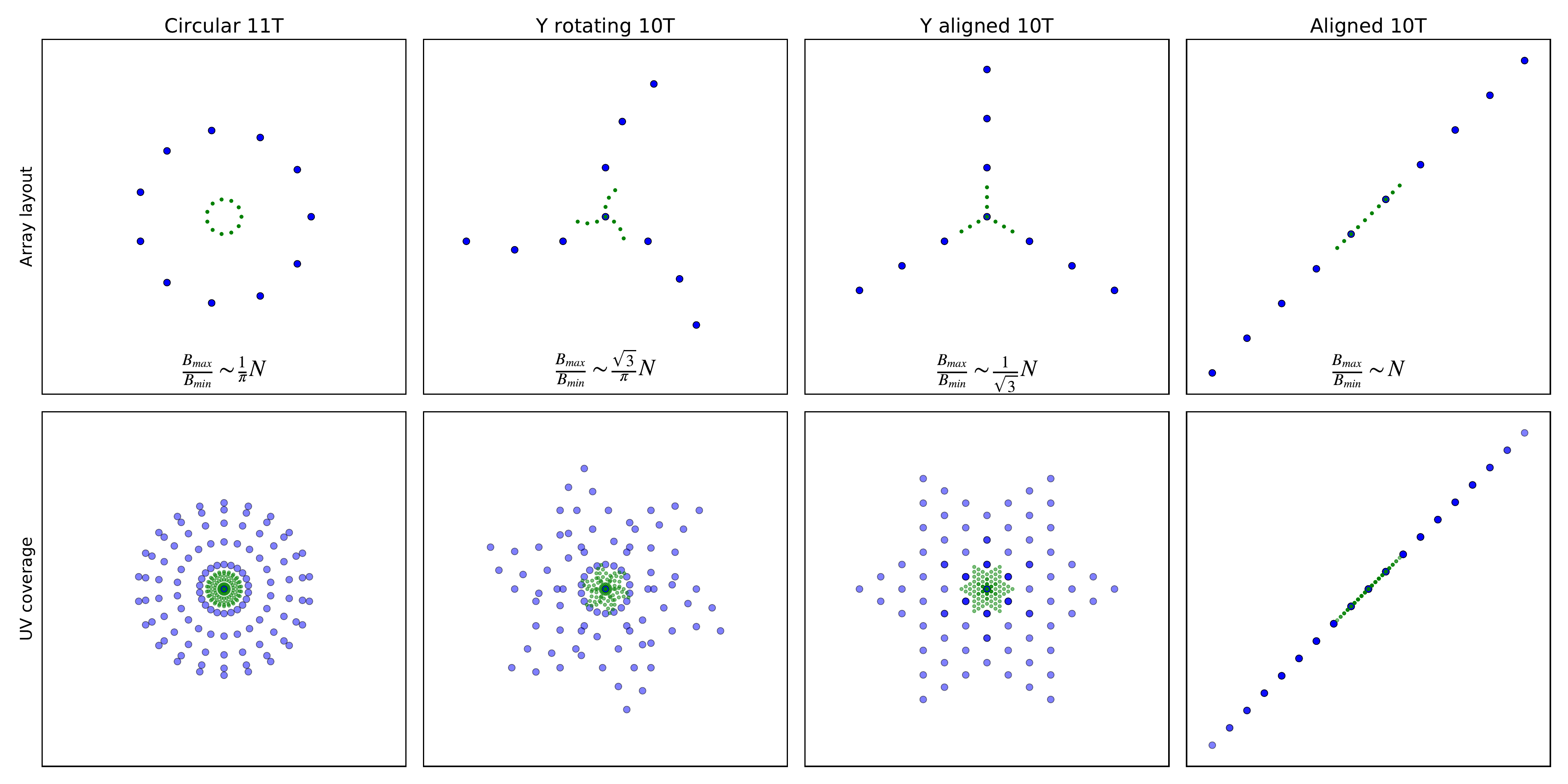}
\caption{The AGILIS concept allows the implementation of many different array configurations, and the adaptation of the array to the size of the observed object. Top: Array configuration. Bottom: Instantaneous (u,~v) coverage at zenith. Left to Right: circular, Y rotating, Y aligned, and linear array configurations, ordered in increasing maximum to minimum baseline ratio ($B_{max}/B_{min}$). Two optimized array scales are presented for observations at a wavelength of \SI{2}{\micro\meter}: $B_{min}=\SI{200}{\meter}$ for a \SI{1}{\milli\arcsecond} object (large blue dots), and $B_{min}=\SI{40}{\meter}$ for a \SI{5}{\milli\arcsecond} object (small green dots).}
\label{Fig:ArrayConfigurations}
\end{figure}


\begin{figure}
\centering
\includegraphics[width=0.5\linewidth]{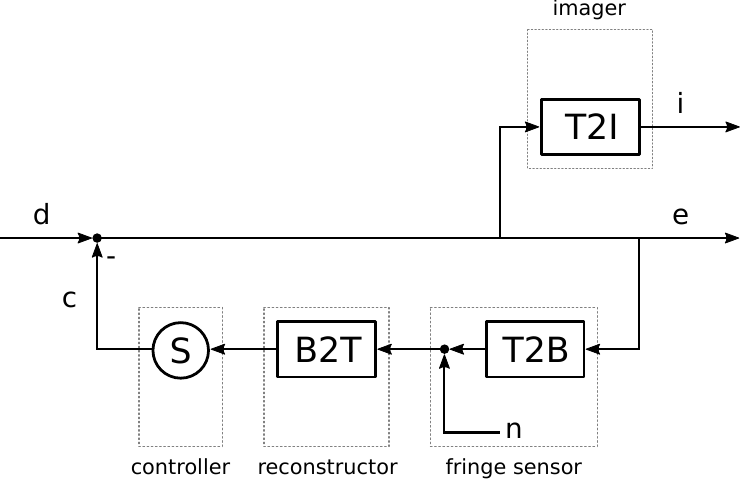}
\caption{Fringe tracking control model. Top: For each telescope, the residual piston error $e$ corresponds to the fringe tracking command $c$ removed from the piston disturbance $d$. The fringe sensor measures differential piston on baselines and introduces noise $n$. The layout of the fringe sensor is contained in the $T2B$ matrix that converts the individual pistons at the level of the telescopes into the differential piston at the level of the fringe sensor baselines.}
\label{Fig:ControlLoop}
\end{figure}

\begin{figure}
\centering
\includegraphics[width=\linewidth]{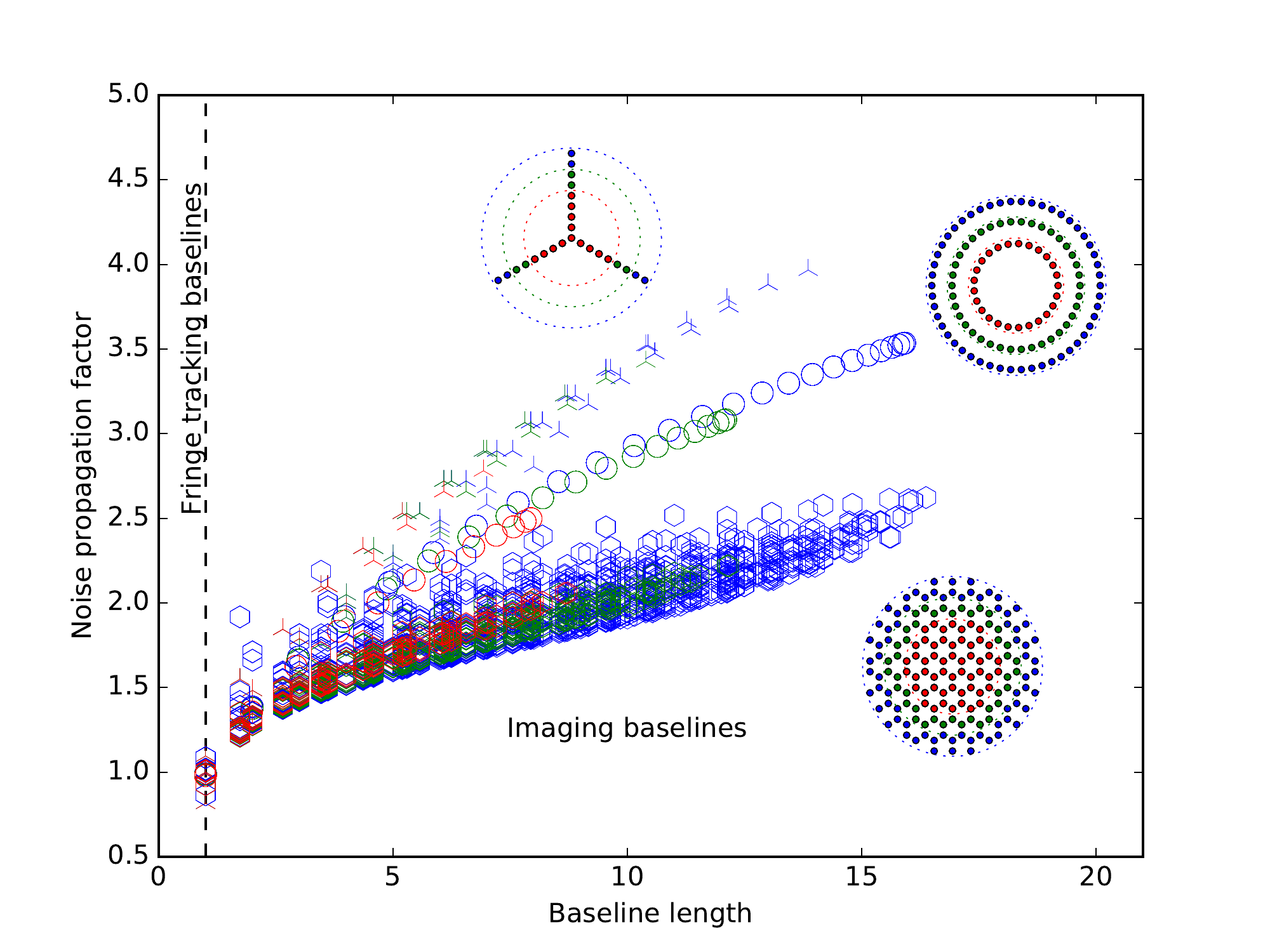}
\caption{Photon noise propagation factor $\sigma_i^2$ (Eq.~\ref{Eq:BaselineNoise}, section \ref{SSec:CophasingArray}), normalized to the photon count per telescope, as a function of the length of the baseline $i$, for three array geometries: Y shape (top), circular (middle), and hexagonal (bottom). The fringe tracking baselines are those at unit baseline length. All arrays are inscribed within a circle of radius of 4 (red), 6 (green), or 8 (blue) unit baseline lengths.}
\label{Fig:NoisePropagation}
\end{figure}

\begin{table*}
\caption{`OHANA Iki transmission error budget, compared to measured transmission. See text for a discussion on the discrepancy.}
\label{tbl:TransmissionErrorBudget}
\centering
\begin{tabular}{ccc}
\hline \hline
Quantity                                                                          & Transmission                          \\
\hline
Photon flux at J=0                                                                & \SI{94673}{\photon\per\milli\second}  \\
\hline
Telescope transmission                                                            & 22.6\%                                \\
Fiber transmission \SI{0.3}{\deci\bel\per\kilo\meter} $\times$ \SI{300}{\meter}   & 97.9\%                                \\
Fiber head Fresnel reflection loss $(95.6\%)^{4}$                                 & 83.5\%                                \\
Delay line transmission	$(98\%)^{10}$                                             & 81.7\%                                \\
Detector quantum efficiency                                                       & 45.0\%                                \\ 
\textbf{Total}                                                                    & \textbf{6.8\%}                        \\
\hline
Expected per telescope                                                            & \SI{6438}{\photon\per\milli\second}   \\
\hline
Measured per telescope                                                            & \SI{283}{\photon\per\milli\second}    \\
\hline
\end{tabular}
\end{table*}


\end{document}